\documentclass[aps,prl,showpacs,amsmath,twocolumn,amssymb,superscriptaddress,letterpaper]{revtex4}
\usepackage{graphicx,color}
\usepackage{amssymb}   % for math
\usepackage{amsmath}
\usepackage{epstopdf}
\usepackage{natbib}
\usepackage{hyperref}
\usepackage{bm}
\usepackage{color}

\begin{document}
\title{Minimal setup for non-Abelian braiding of Majorana zero modes}
\author{Jie Liu}
\affiliation{Department of Applied Physics, School of Science, Xian Jiaotong University, Xian 710049, China}

\author{Wenqin Chen}
\affiliation{Department of Applied Physics, School of Science, Xian Jiaotong University, Xian 710049, China}
\author{Ming Gong}
\affiliation{International Center for Quantum Materials, School of Physics, Peking University, Beijing 100871, China}
\author{Yijia Wu}
\affiliation{International Center for Quantum Materials, School of Physics, Peking University, Beijing 100871, China}
\author{X. C. Xie}
\email{xcxie@pku.edu.cn}
\affiliation{International Center for Quantum Materials, School of Physics, Peking University, Beijing 100871, China}
\affiliation{Beijing Academy of Quantum Information Sciences, Beijing 100193, China}
\affiliation{CAS Center for Excellence in Topological Quantum Computation,
University of Chinese Academy of Sciences, Beijing 100190, China}

\begin{abstract}
Braiding Majorana zero modes (MZMs) is the key procedure toward topological quantum computation. However, the complexity of the braiding manipulation hinders its experimental realization.
Here we propose an experimental setup composing of MZMs and a quantum dot state which can substantially simplify the braiding protocol of MZMs. %With the assistance of a quantum dot,
Such braiding scheme, which corresponds to a specific closed loop in the parameter space, is quite universal and can be realized in various platforms. Moreover, the braiding results can be directly measured and manifested through electric current, which provides a simple and novel way to detect the non-Abelian statistics of MZMs.
\end{abstract}

\maketitle

{\emph {Introduction.}}--- Owing to the charming non-Abelian statistics \cite{kitaev, nayak}, Majorana zero mode (MZM) is deemed as the most promising candidate for topological quantum computation (TQC). To date, topological superconductors (TSCs) that support MZMs have been proposed and realized in various experimental platforms \cite{Fu, sau, fujimoto, sato, alicea2, lut, 2DEG1, 2DEG2, kou, deng, das1,  Marcus, hao, perge, Yaz2, Jia, Fes1, Fes2, Fes3, PJJ1, PJJ2}. Nevertheless, most of the experimental signals so far are based on the zero-bias conductance peaks, which are insufficient to manifest the existence of MZMs \cite{Jie1, brouwer, Aguado, ChunXiao1, Moore, Wimmer, Aguado2,Pan}. The most convincing way demonstrating the presence of MZMs is certainly detecting the signals of its non-Abelian statistics \cite{Ivanov, alicea3, NQP, TQC1}, however, only a few studies have paid attention to this topic thus far \cite{TQC2, TQC3, TQC4}. The main obstruction is that the braiding protocols which have been proposed \cite{MSQ, net1, net2, Yu1,QD1} so far are quite complicated and hard to be realized experimentally.

\begin{figure}
\centering
\includegraphics[width=3.25in]{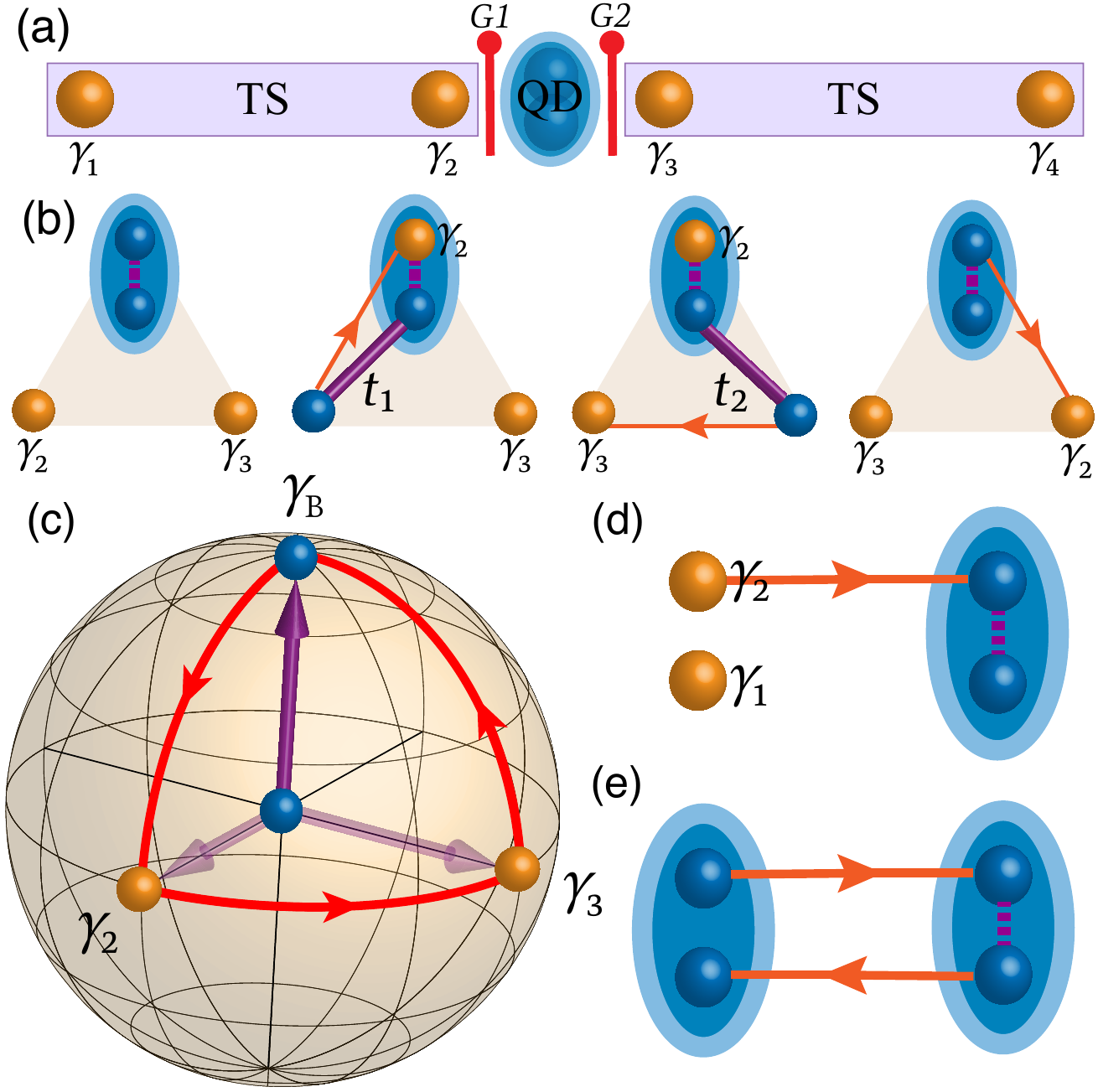}
\caption{
(a) A JJ structure for demonstrating the non-Abelian statistics of MZM.
(b) A minimal sketch consisting of two MZMS and one QD. Through manipulating the coupling strengths, MZMs can be moved and then swapped in the assistance of the QD.
(c) The Bloch sphere of the coupling vector $\vec\delta$. A non-trivial Berry phase is accumulated after a solid angle is spanned by the closed path of $\vec\delta$.
(d) The non-trivial Berry phase induced by tuning the coupling strengths is only picked up by $\gamma_2$ other than by $\gamma_1$. Since two MZMs $\gamma_1$ and $\gamma_2$ together form an ordinary fermionic state as $\psi_1 = \gamma_1+i\gamma_2$, such a fermionic state will not come back to its initial state after braiding.
(e) If the pair of MZMs in (d) are replaced by an ordinary QD state, then although the QD can be decomposed into two MZMs, the non-trivial Berry phase will be picked up by both these two MZMs since they both couple with the other QD with the same strength. Hence, only a global phase will be added to the initial state after braiding.
}
\label{f1}
\end{figure}

The braiding operations usually require a complex 2D network structure to avoid the collision of MZMs during the braiding, where the latter will inevitably happen in a 1D structure \cite{alicea3}. However, as  previous work has demonstrated \cite{ABSJie}, MZM is a Hermitian particle possessing half degree of freedom compared with an ordinary fermion. When an MZM meets an ordinary fermion, it only couples with half of the fermion while leaves another half intact. This means that the moving and swapping of the MZMs through an ordinary fermion could be possible with a proper manipulation.
Thus, the non-Abelian statistics may still be manifested in a quasi-1D structure, though  strictly speaking, the swapping of two MZMs is not allowed in a strict 1D system. Such an idea will greatly facilitate the braiding operation so that it can be realized within the current technology.
The starting point of our proposal follows the traditional braiding protocol of the Y-junction \cite{TQC}, once the simplest proposed platform for carrying out the braiding operations. Moreover, in our proposal, the experimental setup is further simplified by replacing two MZMs with a normal fermionic state. Hence, the platform here is essentially a 1D Josephson junction (JJ) connected through a quantum dot (QD) [Fig. \ref{f1}(a)]. By tuning the coupling strengths between the MZMs and the QD through gate voltages, the MZMs could be moved and swapped after specific manipulations. Moreover, the braiding results are solely determined by the geometric phase corresponding to a specific closed loop in the parameter space, that is time independent and rather robust.

Based on such braiding protocol, we further investigate the electric current signals induced by the braiding of the MZMs. The non-Abelian statistics suggests that the local parity will be reversed during the braiding, corresponding to a local charge transfer process \cite{PTC1, PTC2}. This means that in principle, we can observe a charge transfer from one TSC nanowire to another TSC nanowire. To be explicit, there are two braiding-induced charge transfer processes: the direct electron transition and the Andreev reflection.
%: one process is that an electron in the left nanowire is transmitted to the right nanowire, while the other process is that an electron in the left nanowire is transformed to an hole and be transmitted to the right nanowire.
The total probability of these two processes is exactly equal to the probability of the non-Abelian-braiding-induced parity inversion.
Consequently, although the electric current can not directly manifest the non-Abelian statistics, it does relate to the non-Abelian braiding of MZMs and can be used as an indicator  of the non-Abelian statistics. We further numerically examine all these proposals in a concrete specific model as the semiconductor-superconductor (SS) nanowire, which is one of the most promising TSC platforms. The numerical simulations are fully consistent with the theoretical predictions. This suggests that our proposal is experimentally feasible and hopefully can be performed in the future experimental studies.

{\emph {Simplified setup for non-Abelian braiding of MZMs.}}---Our setup can be straightforwardly constructed by adding an additional QD to the JJ [Fig. \ref{f1}(a)]. Such structure actually has been fabricated in a recent experiment \cite{JJ}. On each TSC nanowire, a pair of MZMs are non-locally distributed at the two ends of the nanowire. Since the MZMs $\gamma_1$ and $\gamma_4$ are far from the interfaces of the JJ due to the length of the TSC nanowire, we can safely neglect them during the braiding. Thus, the simplest Hamiltonian describing such setup is
\begin{equation}
H_s = {2E_d}{d^\dag }d + i({t_1}d + t_1^* {d^\dag }){\gamma_2} + ({t_2}d - t_2^* {d^\dag }){\gamma _3},
\label{eq1}
\end{equation}

\noindent where $d$ is the annihilation operator for the fermioinic state in the QD, and $E_d$ is the on-site energy of this QD state. $\gamma_2$ and $\gamma_3$ are the operators for the MZMs near the interfaces of the JJ, the coupling strengths between the QD and the MZMs $\gamma_2, \gamma_3$ are $t_1=|t_1|e^{i\phi_1/2}$ and $t_2=|t_2|e^{i\phi_2/2}$, respectively, in which the coupling phases $\phi_1$ and $\phi_2$ are the superconducting phase of two TSC.
The crucial feature of this Hamiltonian is that these MZMs couple to the QD in a unique way, because the MZM is Hermitian and possesses half degree of freedom compared with a normal fermionic mode. This feature can be manifested by composing the QD into two intrinsically-fused MZMs $\gamma_A$ and $\gamma_B$ as $d = \frac{1}{2}e^{-i\frac{\phi_1}{2}}(\gamma_A+i\gamma_B)$. In such way, the Hamiltonian $H_s$ can be written in the representation of the Majorana operators as:

\begin{equation}
\begin{aligned}
H_M &= i{E_d}\gamma_A\gamma_B +
i|t_1|\gamma_A{\gamma_2}\\
& + i|t_2| \left[-\gamma_A\sin\left(\frac{\phi_1-\phi_2}{2}\right)+\gamma_B\cos\left(\frac{\phi_1-\phi_2}{2}\right) \right] \gamma_3.
\end{aligned}
\label{eq2}
\end{equation}

\noindent Such Hamiltonian unveils that the coupling between an MZM and the QD can be treated as the coupling between different MZMs. This leads to an idea that the braiding can be performed by individually tuning these two coupling terms.

\begin{figure}
\centering
\includegraphics[width=3.25in]{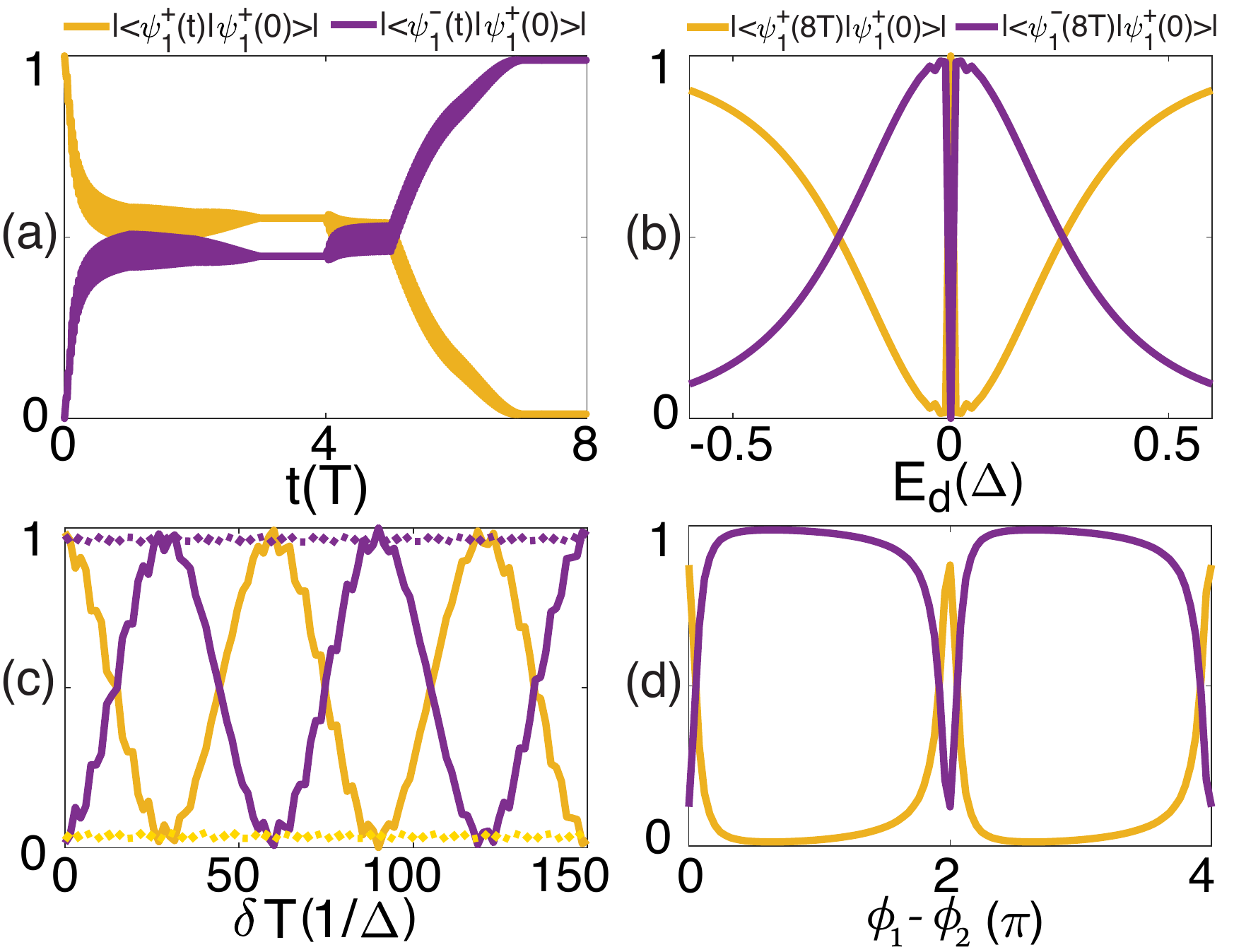}
\caption{
(a) Evolution of the wavefunction $\psi_1^{-}(t)$ during the braiding. After swapping MZMs $\gamma_2$ and $\gamma_3$ twice in succession, $\psi_1^{-}$ evolves into $\psi_1^{+}$ due to the non-Abelian statistics of the MZMs. Here, $E_d=0.04\Delta$, $t_c=0.2\Delta$, and $T=1000/\Delta$.
(b) Braiding results as functions of $E_d$.
(c) Braiding results as functions of the braiding time-cost $T=(1000+\delta T)/\Delta$ with $\phi_1-\phi_2 = \pi+0.05\pi$. Due to the deviation of phase difference $\delta\phi = 0.05\pi$, $\gamma_2$ and $\gamma_3$ will couple with each other and induce a dynamic phase. Such a dynamic phase can be eliminated by the spin-echo technology.
%the non-trivial phase would recovered.
(d) Braiding results as functions of the phase difference $\phi$ when the spin-echo technology is adopted. The non-Abelian braiding properties recovers for almost all the $\phi_1-\phi_2$ values except for $\phi_1-\phi_2=0$.
}
\label{f2}
\end{figure}

The braiding protocol could be greatly simplified by tuning the flux to keep $\phi_1-\phi_2 = \pi$. In this case, the Hamiltonian [Eq. (\ref{eq2})] reduces to an elegant form as $\widetilde{H}_M=i\gamma_A(\vec{\delta}\cdot \vec{\gamma})$, where the Majorana vector $\vec{\gamma} \equiv (\gamma_x(\gamma_2), -\gamma_y(\gamma_3), \gamma_z(\gamma_B))$, and the coupling vector $\vec{\delta} \equiv (|t_1|, |t_2|, E_d)$. Such a mathematical form is in parallel with the traditional Y-junction composed of four MZMs \cite{TQC}, which has been deemed as the simplest model for non-Abelian braiding of MZMs. Accordingly, our proposal follows the braiding scheme adopted in the Y-junction but with a simpler structure and simpler operations [Fig. \ref{f1}(b)].
To initialize the braiding, infinitesimal coupling strengths $|t_1|, |t_2| \ll E_d$ are used to decouple $\gamma_2, \gamma_3$.
After that, the braiding protocol takes four steps (the time-cost for each step is $T$) to swap $\gamma_2$ and $\gamma_3$ spatially. In step 1, $|t_1|$ is increased from $0$ to $t_c$ so that $|t_1|\gg{}E_d$, hence $\gamma_2$ is teleported to the original position of $\gamma_B$. In step 2, $|t_2|$ is increased from $0$ to $t_c$ so that $|t_2| \approx |t_1|\gg E_d$. In step 3, $|t_1|$ is decreased to $0$ so that $|t_1|\ll E_d$ and $\gamma_3$ is teleported to the original position of $\gamma_2$. Finally in step 4, $|t_1|$ is decreased to $0$ so that the spatial positions of $\gamma_2$ and $\gamma_3$ are swapped, although $\vec{\delta}$ comes back to its initial form.

Such a braiding operation, represented by the operator $B(\gamma_i,\gamma_j) = \exp(\frac{\pi}{4}\gamma_i\gamma_j)$, transforms the MZMs as $\gamma_i\rightarrow \gamma_j$ and $\gamma_j\rightarrow -\gamma_i$ \cite{Ivanov}. For two pairs of MZMs whose eigenstates are in the forms of $\psi_j^{\pm}(0) = (\gamma_{2j-1}\pm i\gamma_{2j})/\sqrt{2}$, if $\gamma_2$ and $\gamma_3$ are swapped twice in succession, then the wavefunctions will evolve into $\psi_1^{\pm} (8T) =  (\gamma_1\mp i\gamma_2)/\sqrt{2}=\psi_1^{\mp}(0)$ and $\psi_2^{\pm} (8T) =  (-\gamma_3\pm i\gamma_4)/\sqrt{2}=-\psi_2^{\mp}(0)$.
We calculate the wavefunction evolution during the braiding as $|\psi_j^{\pm}(t)\rangle=U(t)|\psi_j^{\pm}(0)\rangle$, where $U(t) = \hat{T} \exp[i\int_0^{t} \mathrm{d}\tau H_s(\tau)]$ is the time-evolution operator and $\hat{T}$ is the time-ordering operator \cite{Jie2, Jie3, Dirac}. The results confirm that $\psi_{j}^{+}$ evolves into $\psi_{j}^{-}$  after adiabatically swapping $\gamma_2$ and $\gamma_3$ twice in succession [Fig. \ref{f2}(a)], which demonstrates the non-Abelian braiding properties discussed above.

To clearly formulate the braiding process, we can visualize $\vec{\delta}$ as a radius vector in a 3D parameter space in analogy with the Bloch sphere [Fig. \ref{f1}(c)]. During the braiding, $\vec{\delta}$ go along a closed loop on the Bloch sphere that first rotates around the $y$-axis from $(0,0,1)$ to $(1,0,0)$, and then rotates around the $z$-axis from $(1,0,0)$ to $(0,1,0)$. Finally, $\vec{\delta}$ rotates around the $x$-axis back to its start point $(0,0,1)$. Such a closed loop spans an octant of the Bloch sphere and consequently, a geometric phase of $\pi/4$ is accumulated as half of the solid angle of an octant $\Omega_c=\pi/2$.
Furthermore, there is a general relation between the path on the Bloch sphere and the braiding results. The rotation of the coupling vector $\vec\delta$ will bring one MZM, defined as the vector along the polar direction ($\theta$-direction) perpendicular to $\vec\delta$, away from its initial direction towards another MZM which is represented as the vector along the azimuthal direction ($\phi$-direction) perpendicular to $\vec\delta$.
More specifically, such a path corresponding to the swapping of $\gamma_2$ and $\gamma_3$ can be characterized by the braiding operator $U(\gamma_2,\gamma_3) = \exp(\frac{\Omega_c}{2}\gamma_2\gamma_3)$ \cite{S1}, which transforms $\gamma_2$ and $\gamma_3$ as $\gamma_2 \rightarrow \cos(\Omega_c)\gamma_2+\sin(\Omega_c)\gamma_3$, and $\gamma_3 \rightarrow -\sin(\Omega_c)\gamma_2+\cos(\Omega_c)\gamma_3$. Therefore, the braiding results are determined by the solid angle $\Omega_c$ spanned by the loop of $\vec{\delta}$.
In Fig. \ref{f2}(b), we show the weight of $\psi_i^{\pm} (8T)$ on $\psi_i^{\pm} (0)$ after two successive braiding operations, in which $t_1$ and $t_2$ varies in the range of $\left[0 ,t_c \right]$, while $E_d$ remains unchanged. The solid angle is in the general form of $\Omega_c = \arccos(E_d/\sqrt{E_d^2+t_c^2})$, thus the weight of $\psi_2^{\pm} (8T)$ on $\psi_i^{\mp} (0)$ is  $\frac{1-\cos(2\Omega_c)} {2}=\frac{t_c^2}{E_d^2+t_c^2}$. The numerical results in Fig. \ref{f2}(b) are fully consistent with the analysis above except for $E_d=0$.

Now we emphasis that the non-Abelian braiding of MZMs relies on the non-trivial Berry phase as well as the MZM's half fermionic degree of freedom. Since two MZMs $\gamma_1$ and $\gamma_2$ together form a non-local state $\psi = \gamma_1+i\gamma_2$, the non-trivial Berry phase accumulated by tuning the coupling strength (or spatially moving the MZMs) will only be picked up by $\gamma_2$. Thus, the final state cannot return to its original form after braiding. This is completely different from the ordinary fermionic state. If we replace the pair of the MZMs with another QD, then the coupling term between these two QDs is $t_1 d_1^{\dagger} d_1 + h.c.$, where $d_1^{\dagger}$ is the creation operator of the state in the new QD. In the Majorana representation, such coupling term reads as $i|t_1|(\gamma_1\gamma_A-\gamma_B\gamma_2)$, indicating that there are two coupled pairs of MZMs between these two QDs with the same coupling strength [Fig. \ref{f1}(e)]. Then even if a non-trivial Berry phase is accumulated, the initial state will not evolve into another state since both $\gamma_1$ and $\gamma_2$ picks up the same Berry phase. Such an analysis shows the necessity of the half fermionic degree of freedom. Moreover, recent progress \cite{ChunXiao1,  Aguado3} suggests that before the system enters the topological region, an inhomogeneous interface potential could induce a quasi-MZM state comprising of two weakly-coupled MZMs separated in a short distance. Though such state could exhibit non-Abelian braiding properties in specific parameters, such braiding results are extremely fragile %as revealed in the Supplementary Materials \cite{S1}
and can be experimentally distinguished from the case of MZM \cite{S1}.

We have proposed a minimal setup supporting non-Abelian braiding of the MZMs and discussed the underlying physics. However, it should be noted that our proposal requires a $\pi$-junction to facilitate the manipulation. To show what will happen when $\phi_1-\phi_2$ deviates from $\pi$, we calculate the braiding results versus the braiding time-cost $T$ with the phase deviation $\delta\phi=0.05\pi$ [Fig. \ref{f2}(c)]. The braiding results sinusoidally oscillate with $T$, which means that a dynamic phase is induced by the phase deviation. Indeed, since our setup is essentially a JJ with Josephson energy $E_J = \Gamma \cos(\frac{\phi}{2}\gamma_2\gamma_3)$ where $\Gamma$ is the hybridization energy of the junction, an additional dynamic phase will be induced by $\Gamma$. In general, the dynamic phase can be eliminated by the traditional spin-echo technique or parity echo technique, which reverses the sign of the eigenenergy at the middle of the symmetric braiding protocol \cite{GQC}.
Indeed, since this dynamic phase is induced by the Josephson energy, it can be eliminated by reversing the sign of the phase deviation as $\delta \phi=-0.05\pi$ in the last two steps of the braiding so that the non-Abelian geometric phase can be retrieved [the dashed lines in Fig. \ref{f2}(c)]. Figure \ref{f2}(d) exhibits the braiding results as functions of $\phi_1-\phi_2$ when the spin-echo technique is adopted. The non-Abelian braiding results remain intact except for $\phi_1-\phi_2 = 0$, which is quite stable when the spin-echo technique is involved.

{\emph {Non-Abelian-braiding-induced  charge transfer.}}---The detection of the non-Abelian braiding result is another significant problem in the TQC. In the many-body basis $(|0\rangle, \Psi_1^{\dagger}|0\rangle, \Psi_2^{\dagger}|0\rangle, \Psi_2^{\dagger}\Psi_1^{\dagger}|0\rangle)$, $U(\gamma_2,\gamma_3)$
has the matrix form of

\begin{equation}
 \begin{bmatrix}
   \cos(\frac{\Omega_c}{2}) & 0 & 0 &-i\sin(\frac{\Omega_c}{2})\\
   0 & \cos(\frac{\Omega_c}{2}) & -i\sin(\frac{\Omega_c}{2}) & 0\\
   0 & -i\sin(\frac{\Omega_c}{2}) & \cos(\frac{\Omega_c}{2}) & 0\\
   -i\sin(\frac{\Omega_c}{2}) & 0 & 0 & \cos(\frac{\Omega_c}{2}) \\
  \end{bmatrix}.
  \label{eq3}
\end{equation}

\begin{figure}
\centering
\includegraphics[width=3.25in]{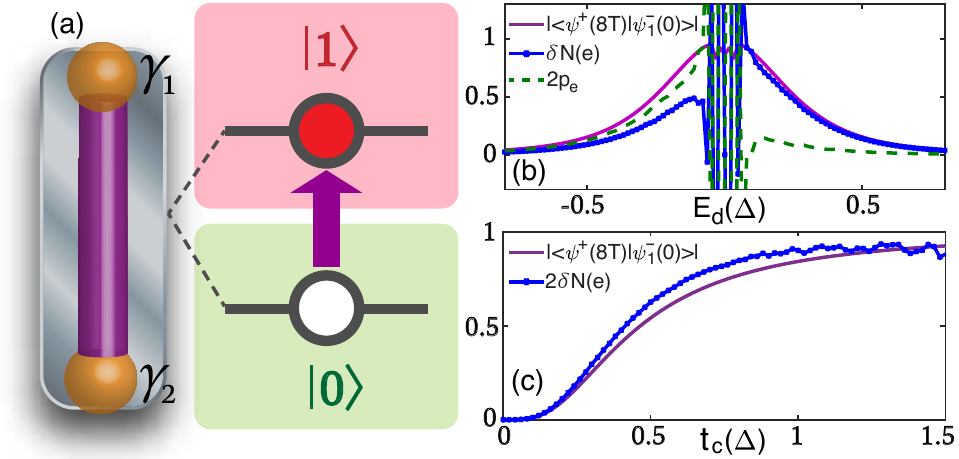}
\caption{
(a) Inversion of the local parity will induce the charge transfer.
(b) Electric-current-induced charge transfer during the braiding (blue line) as a function of $E_d$ with $t_c=0.2\Delta$. The probability of electron tunneling $p_e$ dominates at $E_d<0$ (green dashed line) and the probability of electron tunneling $p_h$ dominates at $E_d>0$. $2(p_e+p_h)$ is equal to the weight of $|\langle \psi_1^{\mp}(0)|\psi_1^{\pm} (8T)\rangle|$ (purple line). Since $\delta N=(p_e+2p_h)e$,  the charge transfer is about $t_c^2/2(E_d^2+t_c^2)$ at $E_d<0$ and about $t_c^2/({E_d^2+t_c^2})$ at $E_d>0$. Resonant charge transfer happens near $E_d=0$ in which both these two processes matter.
(c) The charge transfer as a function of $t_c$ with $E_d=-0.6\Delta$. The charge transfer is about half of $|\langle \psi_1^{\mp}(0)|\psi_1^{\pm} (8T)\rangle|$.
}
\label{f3}
\end{figure}

\noindent The non-zero off-diagonal elements in the braiding matrix demonstrate that an initial state $a|0\rangle+b\Psi_2^{\dagger}\Psi_1^{\dagger}|0\rangle$ will evolve into $\left[a \cos(\frac{\Omega_c}{2})-ib\sin(\frac{\Omega_c}{2}) \right] |0\rangle+ \left[b \cos(\frac{\Omega_c}{2})-ia\sin(\frac{\Omega_c}{2})\right]\Psi_2^{\dagger}\Psi_1^{\dagger}|0\rangle$ after braiding. Since the inversion of the local parity of $\Psi_1$ or $\Psi_2$ corresponds to a charge transfer from the left nanowire to the right one [Fig. \ref{f3}(a)], the charge transfer in an isolated system is equal to $(|a|^2-|b|^2)\sin^2(\Omega/2)$ and can be detected via charge sensing \cite{PTC1, PTC2}. Remarkably, we find that the electric current can also manifest the non-Abelian braiding in an open system. %Since the charge transfer is induced by the non-Abelian statistics of MZMs.
The charge transfer here can be calculated through the integration of the time-dependent current as $\delta N = \int_0^{8T}\langle \hat{J}(t) \rangle \mathrm{d}t$, where the current  follows the definition of \cite{Lei}

\begin{equation}
 \langle \hat{J}(t) \rangle = -\frac{i}{2}
 \left[ \langle \psi_{L}(t)|H_c(t)|\psi_{R}(t) \rangle -h.c. \right].
 \label{eq4}
 \end{equation}

\noindent Here, $|\psi_{L}(t)\rangle$ is the wavefunction at the right end of the left nanowire at time $t$, while $|\psi_{R}(t) \rangle$ is the wavefunction at the left end of the right nanowire at time $t$. Since $H_c(t) = i({t_1}d + t_1^* {d^\dag }){\gamma_2}$, the current is contributed from two possible processes. The first process (with probability denoted by $p_e$) is that an electron in the left nanowire directly transmits into the right nanowire. The other one is an electron in the left nanowire transforms into a hole and then transmits into the right nanowire (Andreev reflection, whose probability is denoted as $p_h$).
Since a positive $E_d$ indicates a barrier for electron transmission, we can see that $p_h$ dominates for $E_d>0$ while $p_e$ dominates for $E_d<0$ [the dashed line in Fig. \ref{f3}(b)]. %They are symmetric about $E_d=0$
Moreover, it can be proved that $2(p_e+p_h)=\frac{t_c^2}{E_d^2+t_c^2}$, which is exactly the population of the charge occupied after braiding $|\langle \psi_i^{\pm}(8T) | \psi_i^{\mp}(0) \rangle|$.
In comparison, the actual charge transfer is $\delta N = p_e+2p_h$ [Fig. \ref{f3}(b)], hence the charge transfer is highly related to the non-Abelian geometric phase that $\delta N \approx |\langle \psi_i^{\pm}(8T) | \psi_i^{\mp}(0) \rangle|/2$ for $E_d<0$, while $\delta N \approx |\langle \psi_i^{\pm}(8T) | \psi_i^{\mp}(0) \rangle|$ for $E_d>0$. Figure \ref{f3}(c) shows the charge transfer as a function of $t_c$ with $E_d=-0.6\Delta$, in which $\delta N \approx |\langle \psi_i^{\pm}(8T) | \psi_i^{\mp}(0) \rangle|/2$ in a large region of $t_c$. These results indicate that the non-Abelian statistics may be directly detected through the electric current.

\begin{figure}
\centering
\includegraphics[width=3.25in]{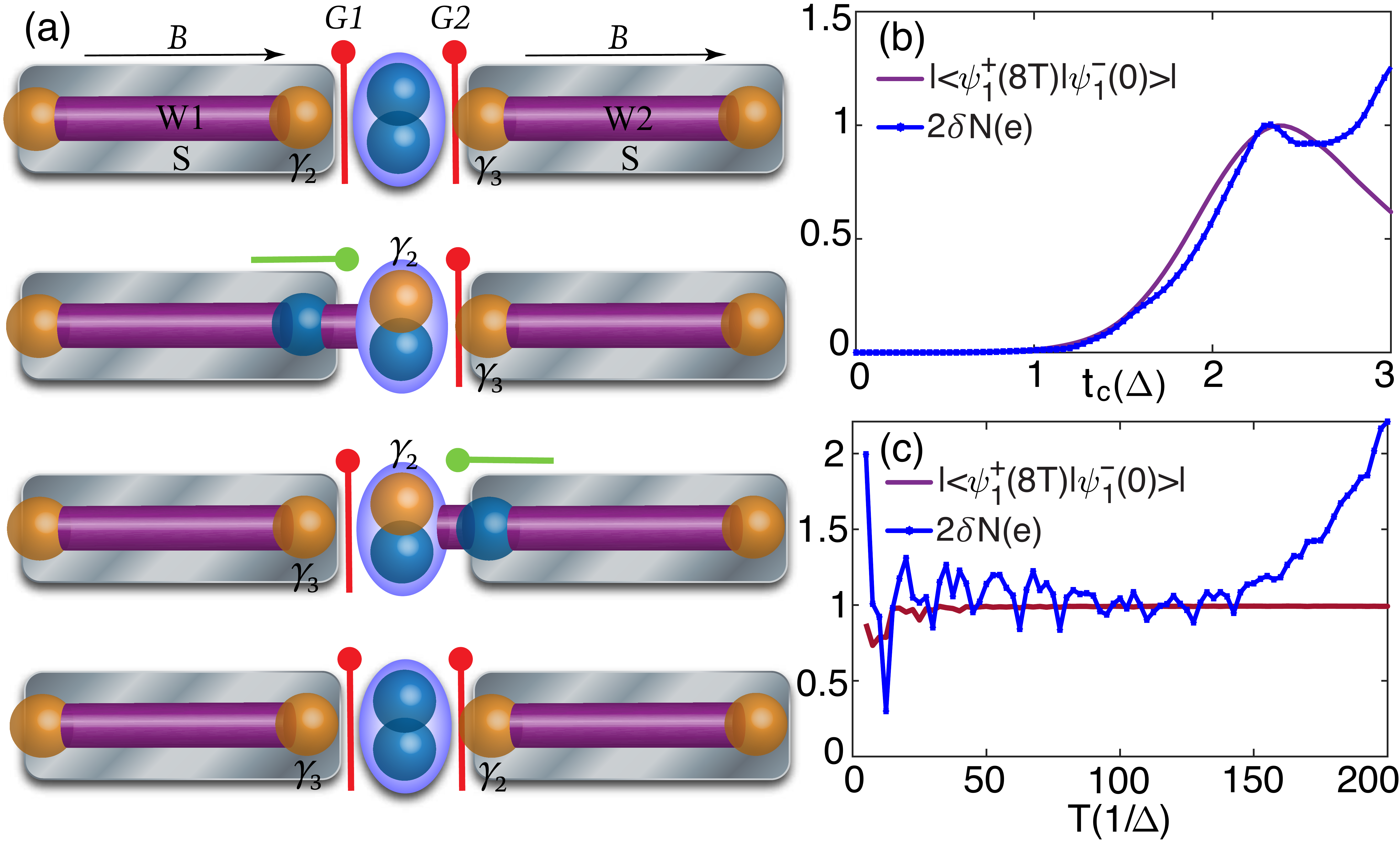}
\caption{
(a) Braiding operation steps in the SS nanowire system.
(b) The braiding result (purple line) and the charge transfer (blue line) in the SS nanowire versus $t_c$
with $E_d=-0.6\Delta$ and braiding time-cost $T=80/\Delta$.
(c) The braiding result in the SS nanowire as a function of $T$ with $E_d=-0.4\Delta$ and $t_c=1.5\Delta$. The charge transfer (blue line) significantly increases when $T > 150/\Delta$ due to the finite size effect.}
\label{f4}
\end{figure}

{\emph {Numerical simulation in the semiconductor superconducting nanowires.}}---We have demonstrated that the non-Abelian braiding of MZMs can be well performed through the protocol we proposed. Here, we would like to numerically examine it in a specific system. Since the semiconductor superconducting (SS) nanowire is deemed as the most promising platform for TQC, we numerically simulate the non-Abelian braiding in a system composed of two SS nanowires [Fig. \ref{f4}(a)], in which the coupling strengths between the MZMs and the QD can be manipulated through two voltage gates located near the intersection of the nanowires and the QD. The whole braiding process is quite simple and can be performed by first turning off the gate voltages in $G1$ and $G2$ simultaneously, and then turning on $G1$ and $G2$ in succession [Fig. \ref{f4}(a)]. The wavefunction evolution and the charge transfer during the braiding can be numerically demonstrated as shown in Fig. \ref{f4}(b), (c). We can see that the weight of of $\psi_i^{\pm} (8T)$ on $\psi_i^{\mp} (0)$ is still $\frac{1-\cos(2\Omega_c)} {2}=\frac{t_c^2}{E_d^2+t_c^2}$, though the on-site energy of the QD state shifts due to the influence of background. Besides, the electron tunneling dominates due to the negative shift of the on-site energy of the QD state. Finally, the charge transfer significantly increases when $T > 150/\Delta$ due to the finite size effect of the nanowire \cite{S1}, implying that its non-Abelian braiding properties will be better exhibited in a longer nanowire.

{\emph {Conclusion.}}---With the assistance of a QD, we show that the non-Abelian braiding operations can be manifested in a simple 1D structure. Such braiding operation is quite simple and feasible in the state-of-art technology. We want to stress that though we take the SS nanowire as a demonstration, our minimal model is quite general. Such proposed minimal model can be realized in various TSC platforms such like ferromagnetic atomic chains and vortices on iron-based TSC.
Moreover, in the SS nanowire system we have demonstrated, in addition to tuning the coupling strengths between the MZMs and the QD, the non-Abelian braiding operations corresponding to the closed loop on the Bloch sphere in Fig. \ref{f1}(c) can also be realized through manipulating the on-site energy of the QD state $E_d$.

{\emph{Acknowledgement.}}--- We acknowledge Lei Wang for the numerical support.
This work is financially supported by NSFC (Grants No. 11974271) and NBRPC (Grants No. 2017YFA0303301, and No. 2019YFA0308403).

\end{document}